\documentclass[aps,pra,reprint,groupedaddress]{revtex4-1}
\usepackage{graphicx}% Include figure files
\usepackage{bm}% bold math
\usepackage{amsmath}

\usepackage[dvipdfm,colorlinks,linkcolor=blue, urlcolor=blue, anchorcolor=blue, citecolor=blue]{hyperref}

\begin{document}
%\preprint{}
\title{Highly efficient charging and discharging of three-level quantum batteries through shortcuts to adiabaticity}


\author{Fu-Quan Dou}
\email[]{doufq@nwnu.edu.cn}
%\thanks{}
%\altaffiliation{doufq@nwnu.edu.cn}
%\affiliation{College of Physics and Electronic Engineering, Northwest Normal University, Lanzhou, 730070, China}
\author{Yuan-Jin Wang}%
\affiliation{
College of Physics and Electronic Engineering, Northwest Normal University,
Lanzhou, 730070, China%\\This line break forced with \textbackslash\textbackslash
}%
\author{Jian-An Sun}%
\affiliation{
College of Physics and Electronic Engineering, Northwest Normal University,
Lanzhou, 730070, China%\\This line break forced with \textbackslash\textbackslash
}%

%\date{\today}

\begin{abstract}
Quantum batteries are energy storage devices that satisfy quantum mechanical principles. %mobility and portability. a significant field of research
 How to improve the battery's performance such as stored energy and power is a crucial element in the quantum battery. Here, we investigate the charging and discharging dynamics of a three-level counterdiabatic stimulated Raman adiabatic passage quantum battery via shortcuts to adiabaticity, which can compensate for undesired transitions to realize a fast adiabatic evolution through the application of an additional control field to an initial Hamiltonian. %We show a highly efficient charging and discharging in the three-level counterdiabatic stimulated Raman adiabatic passage quantum battery.
  The scheme can significantly speed up the charging and discharging processes of a three-level quantum battery and obtain more stored energy and higher power compared with the original stimulated Raman adiabatic passage. We explore the effect of both the amplitude and the delay time of driving fields on the performances of the quantum battery. Possible experimental implementation in superconducting circuit and nitrogen-vacancy center is also discussed. %we apply the technology shortcuts to adiabaticity to a three-level quantum battery charged or discharged through the widespread stimulated raman adiabatic passage and prove that this method can improve the power of quantum battery. More specifically, we realizes shortcuts to adiabaticity through the application of an additional control field  to compensate undesired transitions, then the charging or discharging process of the quantum battery is numerically simulated to confirm the accelerated utility of shortcuts to adiabaticity. Meanwhile, we also explore the relationship between the performance of the quantum battery and  the amplitude or the delay time of the driving fields. This new charging or discharging protocol can be implemented in superconducting circuit and N-V centre. quantum mechanical systems that are able to store energy

\end{abstract}

%\keywords{}

\maketitle

\section{Introduction}
Quantum batteries are quantum mechanical systems which are able to store or release energy and satisfy the requirement of miniature battery \cite{Campaioli2018, bhattacharjee2020quantum}. Currently widely used batteries are electrochemical devices \cite{quach2020using} that store energy into chemical energy and convert it into electrical energy to drive machines to work. As many electronic devices continue to be miniaturized to provide flexibility and portability  \cite{sen2019local} and quantum technologies continue to make progress, it has become a prevalent topic to construct completely new energy storage devices called quantum batteries. % which transcend traditional electrochemical batteries with quantum systems.
  They are based on quantum thermodynamics, which are fundamentally different from and transcend the traditional electrochemical batteries \cite{quach2020using}. The influences of quantum phenomena \cite{kamin2020entanglement, PhysRevLett.125.040601,Carrega_2020,PhysRevLett.122.210601,PhysRevA.100.043833, PhysRevLett.111.240401,tabesh2020environmentmediated,PhysRevLett.118.150601,PhysRevResearch.2.013095} such as coherence and entanglement on quantum batteries must be taken into account.

Since the introduction of quantum batteries in 2013 \cite{PhysRevE.87.042123}, quantum phenomena have been used to improve the performances of the quantum batteries in various models \cite{PhysRevLett.124.130601, Kamin_2020}, such as quantum cavity \cite{PhysRevLett.120.117702, zhang2018enhanced, PhysRevA.100.043833,PhysRevLett.122.047702,Binder_2015,10.1088/1367-2630/ab91fc,
PhysRevResearch.2.023113,mohan2020reverse,niedenzu2018quantum,PhysRevResearch.2.023095}, spin chain \cite{PhysRevA.97.022106,PhysRevA.101.032115,PhysRevA.103.033715,PhysRevB.100.115142,
zakavati2020bounds,ghosh2020fast,PhysRevE.101.062114}, Sachdev-Ye-Kitaev model \cite{rossini2019quantum,rosa2019ultra} and quantum oscillators \cite{PhysRevE.99.052106, PhysRevB.98.205423, PhysRevB.99.205437, Chen_2020}, showing quantum advantages. In the quantum cavity, a Dicke quantum battery is constructed by coupling $N$ two-level systems with a single photonic mode in a cavity \cite{zhang2018enhanced}. Interestingly, this collective charging protocol increases the charging power of a quantum battery by $\sqrt{N}$ times than the parallel one that assigns a single optical cavity to each two-level system \cite{PhysRevLett.120.117702}. In fact, this collective quantum advantage that enhances the charging power of quantum batteries is widespread \cite{PhysRevLett.118.150601}. Moreover, the charging power of a many-body quantum battery is further enhanced by considering the spin-spin interaction in the spin chain \cite{PhysRevA.97.022106}. The exactly solvable Sachdev-Ye-Kitaev model is also a suitable cell for describing quantum batteries, which has shown that the presence of non-local correlations can significantly inhibit the fluctuations of the average energy stored in quantum batteries \cite{rosa2019ultra}. The quantum battery in quantum oscillators model consists of $N$ two-level atoms, and is charged through a harmonic charging field, which is significantly different from a static charging field. In the absence of interatomic interaction, it can be fully charged by appropriately adjusting driving frequency. The fully charging power can be further enhanced when repulsive interaction exists between large $N$ atoms \cite{PhysRevE.99.052106}.

Recently, based on stimulated Raman adiabatic passage (STIRAP) technique, one propose a three-level quantum battery \cite{PhysRevE.100.032107,Dou_2020}, which is a three-level system driven by two external optical or microwave fields that realizes the energy transfer between the ground state and the maximum excited state along the so-called dark state \cite{RevModPhys.89.015006,RevModPhys.70.1003,Shore:s}. The characteristics of three-level systems that base on STIRAP to achieve quantum transitions have been studied in detail. For example, the loss due to spontaneous emission of the intermediate state is negligible, and the process is relatively insensitive to experimental imperfections in pulse intensity, shape and timing \cite{Shore:s}. These characteristics suppress undesired spontaneous discharging, and greatly enhance the stability and robustness of a three-level quantum battery charged by STIRAP \cite{PhysRevE.100.032107}. On that basis, it is shown that when an appropriate third driving field is applied to STIRAP, the transitions between the three quantum states can form a closed loop, which can effectively improve the charging power of the quantum battery \cite{Dou_2020}. However, in order to maintain the system as closely as possible in the dark state, in general, it requires rather long time to satisfy adiabaticity.

The protocol of shortcuts to adiabaticity (STA) \cite{PhysRevLett.105.123003,RevModPhys.91.045001} is capable of accelerating the evolution in STIRAP and can be achieved by implementing an auxiliary field to suppress the nonadiabatic excitation in STIRAP. This method is also named counterdiabatic, transitionless or superadiabatic driving \cite{PhysRevA.96.013431,dou2017high,Berry_2009,demirplak2003adiabatic,del2013shortcuts,PhysRevA.89.033419}. It has been widely used in atomic, molecular, optical and other fields \cite{RevModPhys.91.045001,PhysRevLett.122.173202}. Experimentally, it has been implemented in superconducting circuit \cite{vepsalainen2019superadiabatic}, nitrogen-vacancy (N-V) centre \cite{barfuss2018phase,PhysRevLett.122.090502,PhysRevLett.110.240501}, and cold atoms \cite{Schaff_2011,du2016experimental} etc. Very recently, the transitionless driven quantum batteries, composed by a set of independent three-qubit cells are proposed and the energy resources can be further optimized by implementing the transitionless technique \cite{moraes2020charging}.

In this paper, we investigate the characteristics of a three-level quantum battery when it is charged and discharged through STA. More specifically, based on counterdiabatic driving, we require an additional control field to ensure adiabaticity and form a three-level counterdiabatic STIRAP (cdSTIRAP) quantum battery. We are primarily concerned with the ergotropy which is the maximal energy can be extracted from the quantum battery to perform thermodynamic work \cite{Allahverdyan_2004,PhysRevE.102.022106,niedenzu2018quantum,
akmak2020ergotropy,akmak2020ergotropy,ito2020collectively,tacchino2020non} and power of the quantum battery. We consider the counterdiabatic protocol will enhance the charging and discharging power. To prove it, we study the properties of the cdSTIRAP quantum battery %by means of numerical simulations
 and compare them with STIRAP case. Meanwhile, we also simulate the effects of the amplitude and the delay time of driving fields on the charging process \cite{mohan2020reverse}. Finally, we discuss the feasible schemes of our cdSTIRAP quantum battery in experiments.

The paper is organized as follows. In Sec. \uppercase\expandafter{\ref{2}}, we show a three-level quantum battery and charging (discharging) protocols. %At the same time,
 The Hamiltonian of the quantum battery involved in the corresponding protocol is also introduced. Next in Sec. \uppercase\expandafter{\ref{3}}, we analyze the characteristics of the two charging protocols and calculated the corresponding maximum charging energy and average power. %by numerical simulation.
 Then, the discharging processes are analyzed through STA in Sec. \uppercase\expandafter{\ref{4}}. We explore the feasible schemes of our cdSTIRAP quantum batteries in experiments in Sec. \uppercase\expandafter{\ref{5}}. Finally, a brief summary is reported in Sec. \uppercase\expandafter{\ref{6}}.

\section{\label{2} Model}%
\begin{figure}
	\includegraphics[width=0.5\textwidth]{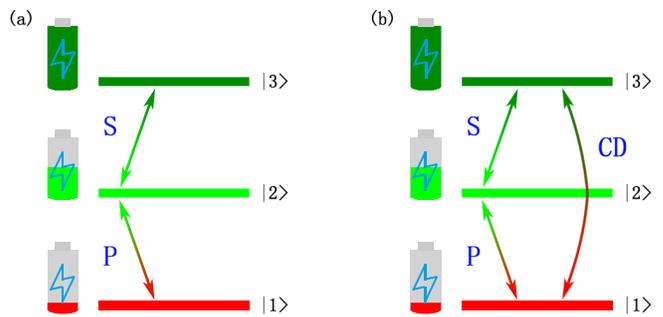}
	\caption{\label{Fig1} Three-level quantum battery configuration and charging (discharging) schemes. (a)\label{1.a} STIRAP charging (discharging) protocol. The ground state $\left|1\right\rangle$, the first excited state $\left|2\right\rangle$ and the second excited $\left|3\right\rangle$ are colored red, light green and dark green respectively, representing the bare (discharged) battery, partially charged battery and fully charged battery in turn.  The charging of the quantum battery is realized by the transitions between quantum states. We utilize P pulse to drive $|1\rangle-|2\rangle$ transition and S pulse to realize $|2\rangle-|3\rangle$ transition. (b)\label{1.b} cdSTIRAP charging (discharging) protocol. On the basis of (a), to compensate for nonadiabatic transitions, applying additional field CD to control $|1\rangle-|3\rangle$ transition.}
\end{figure}

In this section, we first briefly introduce the three-level quantum battery. Suppose that the energy eigenstate of a three-level system is $\left|\varepsilon_{i}\right\rangle (i=1,2,3)$ and the corresponding energy is $\varepsilon_{i} (i=1,2,3)$. The ground state $\left|\varepsilon_{1}\right\rangle$ denotes a discharged battery. When the system is in the first excited state $\left|\varepsilon_{2}\right\rangle$, it is similar to a partially charged quantum battery. And the system requires to be in the second excited state $\left|\varepsilon_{3}\right\rangle$ when the quantum battery is fully charged (see Fig.\ref{Fig1}).

As shown in Fig. \hyperref[1.a]{1(a)}, the STIRAP charging or discharging protocol involves a P pulse on $|1\rangle-|2\rangle$ transition and a S pulse on $|2\rangle-|3\rangle$ transition. And the complete Hamiltonian under a time-dependent interaction picture reads \cite{PhysRevE.100.032107, PhysRevLett.122.173202, barfuss2018phase}
\begin{equation}
H_{1}=\frac{\hbar}{2}\left[\begin{array}{ccc}
0 & \Omega_{p} & 0 \\
\Omega_{p} & 0 & \Omega_{s} \\
0 & \Omega_{s} & 0
\end{array}\right],
\end{equation}
where $\hbar$ is the reduced Planck constant $(\hbar=1)$, $ \Omega_{p}$, $\Omega_{s}$ respectively the amplitudes of the corresponding driving fields.

The three instantaneous eigenstates \cite{PhysRevLett.122.090502} of Hamiltonian $(1)$ are
\begin{equation}\begin{aligned}
&\left|-\right\rangle=\frac{\sin \theta|1\rangle-|2\rangle+\cos \theta|3\rangle}{\sqrt{2}}, \\
&\left|0\right\rangle=\cos \theta|1\rangle-\sin \theta|3\rangle,\\
&\left|+\right\rangle=\frac{\sin \theta|1\rangle+|2\rangle+\cos \theta|3\rangle}{\sqrt{2}},
\end{aligned}\end{equation}
and the corresponding eigenvalues are $\lambda_{-}=-\Omega, \lambda_{0}=0, \lambda_{+}=\Omega$, respectively. Here $\Omega=\sqrt{\Omega_{p}^{2}+\Omega_{s}^{2}}$, and $\tan \theta(t)=\Omega_{p}(t) / \Omega_{s}(t)$. The intermediate state $\left|0\right\rangle$ is called the dark state, which connects the $|1\rangle$ and $|3\rangle$ states without need of state $|2\rangle$.

Fig. \hyperref[1.b]{1(b)} show the cdSTIRAP protocol. The only difference from STIRAP protocol is that an additional driving field CD is applied to compensate for the nonadiabatic excitation. Thus, the complete interaction Hamiltonian becomes
\begin{equation}
H_{2}=H_{1}+H_{cd},
\end{equation}
with
\begin{equation}
H_{cd}=\frac{\hbar}{2}\left[\begin{array}{ccc}
0 & 0 & \Omega_{cd} e^{i \phi} \\
0 & 0 & 0 \\
\Omega_{cd} e^{-i \phi} &0 & 0
\end{array}\right].
\end{equation}
Here, $\Omega_{cd}$ is the amplitude of the CD field and phase $\phi=\pi/2$.

For the purpose to obtain exact magnitude of $\Omega_{cd}$ and maintain the adiabatic evolution of the system, we convert the Hamiltonian $H_{2}$ into a new matrix with three eigenstates in $(2)$ as the base vector. The transformation matrix \cite{PhysRevLett.122.173202} is
\begin{equation}
U=\left[\begin{array}{ccc}
\frac{\sin \theta}{\sqrt{2}} & -\frac{1}{\sqrt{2}} & \frac{\cos \theta}{\sqrt{2}} \\
\cos \theta & 0 & -\sin \theta \\
 \frac{\sin \theta}{\sqrt{2}} & \frac{1}{\sqrt{2}} & \frac{\cos \theta}{\sqrt{2}}
\end{array}\right].
\end{equation}
Then, we make use of $H^{'}_{2}$ to represent the new matrix, which satisfies
\begin{equation}
H^{'}_{2}=UH_{2}U^{-1}+\mathrm{i} \frac{\mathrm{d} U}{\mathrm{d} t} U^{-1}.
\end{equation}
Apparently,
\begin{equation}
H^{'}_{2}=\left[\begin{array}{ccc}
-\frac{1}{2} \Omega&\frac{i}{\sqrt{2}}\left(\dot{\theta} - \frac{\Omega_{cd}}{2} \right) & 0 \\
\frac{-i}{\sqrt{2}}\left(\dot{\theta} - \frac{\Omega_{cd}}{2} \right) & 0 & \frac{-i}{\sqrt{2}}\left(\dot{\theta} - \frac{\Omega_{cd}}{2} \right) \\
0 & \frac{i}{\sqrt{2}}\left(\dot{\theta} - \frac{\Omega_{cd}}{2} \right) & \frac{1}{2} \Omega
\end{array}\right].
\end{equation}
Note that when we make
\begin{equation}
\Omega_{cd}(t)=2 \dot{\theta}(t),
\end{equation}
the off-diagonal elements in $H^{'}_{2}$ will vanish \cite{PhysRevLett.122.173202, vepsalainen2019superadiabatic,petiziol2020optimized}. It means that the system will be locked in the dark state during adiabatic evolution, i.e., the three-level quantum battery can be charged or discharge through STA.

If we use density matrix $\rho_{int}$ to describe the quantum state of the system at $t$ moment, i.e.,
\begin{equation}
\rho_{int}=\sum_{n=1}^{3} r_{n}\left|r_{n}\right\rangle\left\langle r_{n}\right|, \quad r_{1} \ge r_{2} \ge r_{3},
\end{equation}
the dynamics \cite{Allahverdyan_2004} can be written as
\begin{equation}
\dot{\rho}_{\mathrm{int}}(t)=\frac{1}{i \hbar}\left[H_{1,2}(t), \rho_{\mathrm{int}}(t)\right].
\end{equation}
And the total energy can be simply described as \cite{PhysRevB.100.115142}
\begin{equation}
E(t)=\sum_{n} r_{n} \varepsilon_{n}, n=1,2,3.
\end{equation}

During charging and discharging, the ergotropy in quantum battery is
\begin{equation}
C(t)=E(t)-\varepsilon_{1}.
\end{equation}
The charging power defines
\begin{equation}
\mathrm{P}(\mathrm{t})=\frac{C(t)}{t},
\end{equation}
while the discharging power is
\begin{equation}
\mathrm{P}_{d}(\mathrm{t})=\frac{\varepsilon_{3}-E(t)}{t}.
\end{equation}

\section{\label{3}Charging Dynamics}%	
We now study the charging characteristics of quantum battery via cdSTIRAP protocol. In following analyses, we consider the driving fields are Gaussian pulses, of the form
\begin{eqnarray}
\displaystyle \Omega_{p}(t)=\Omega_{0} e^{-\left(\frac{t-\tau}{T}\right)^{2}}, \Omega_{s}(t)=\Omega_{0} e^{-\left(\frac{t+\tau}{T}\right)^{2}},
\end{eqnarray}
with delay $\tau$, peak value $\Omega_{0}$ and width $T$.
The additional driving field CD is realized by modulating
\begin{equation}
\Omega_{cd}(t)=\frac{4\tau}{T^{2}} \mathrm{sech}\left(\frac{4 \tau}{T^{2}}t\right).
\end{equation}

\begin{figure}
	\includegraphics[width=0.5\textwidth]{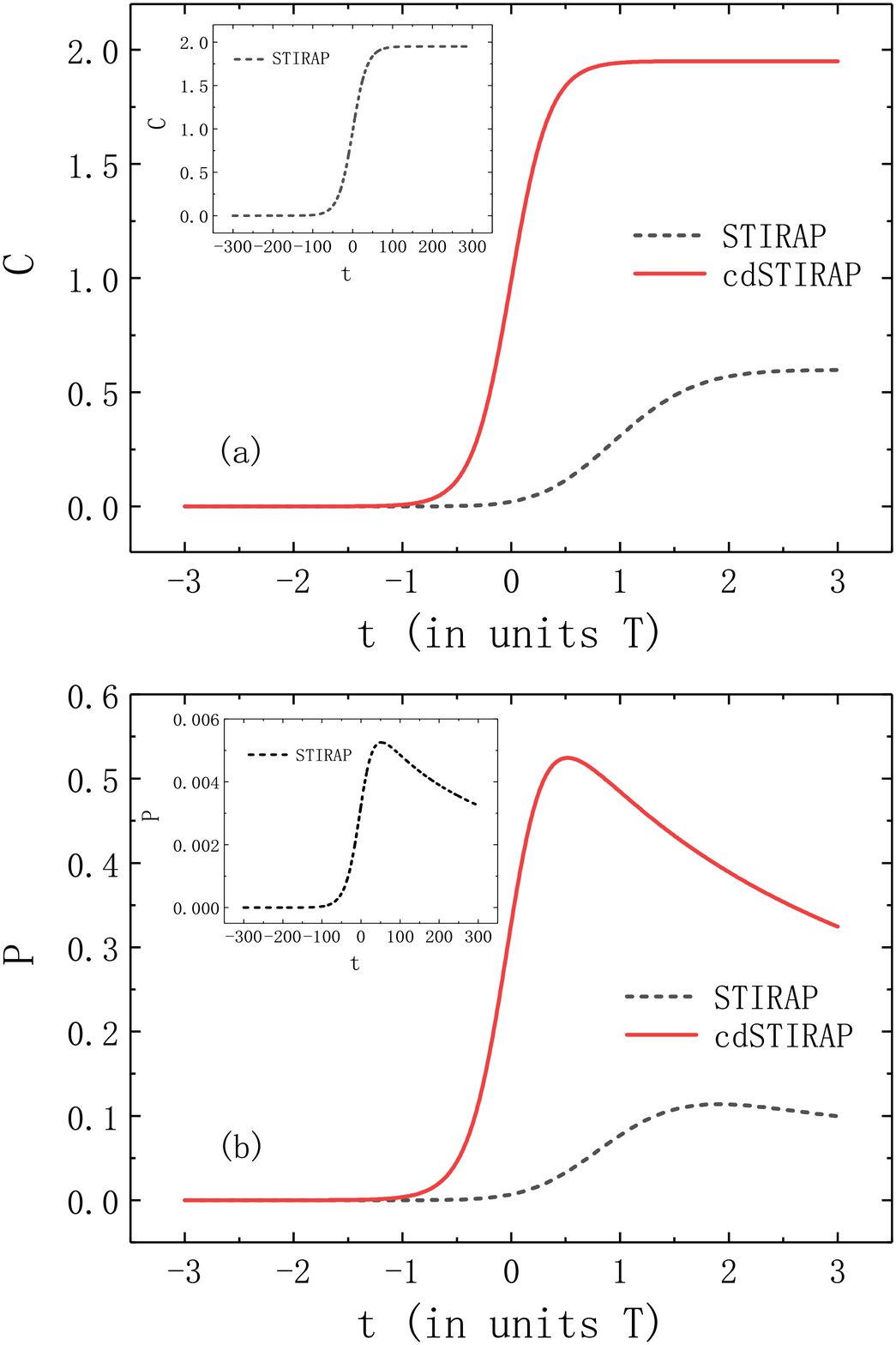}
	\caption{\label{Fig2} Time evolution of ergotropy (a) and power (b) during charging. The red solid line corresponds to cdSTIRAP charging protocol and the black dashed line represents the STIRAP charging protocol. The parameters selected for these data are width $T=1$, peak value $\Omega_{0}=1$, and delay $\tau=0.7T$. (Inset) Same as in the main panel but for $T=100$, representing standard STIRAP charging protocol. We chose the energy spectrum of the three-level system as $\varepsilon_{1}=0, \varepsilon_{2}=1$ and $\varepsilon_{3}=1.95$. }
\end{figure}

In all following calculations, for simplicity, we choose the energy spectrum of the three-level system as $\varepsilon_{1}=0, \varepsilon_{2}=1$ and $\varepsilon_{3}=1.95$. Fig. \ref{Fig2} shows the variation characteristics of ergotropy and power over time under specific parameters. The red solid line indicates the adiabatic evolution in cdSTIRAP charging protocol.
For comparison purposes, the STIRAP charging protocol is also depicted with a black dashed line. As shown in Fig. \hyperref[Fig2]{2(a)}, the cdSTIRAP charging protocol realize the complete and stable charging of the quantum battery in a very short time. During the entire charging process, the ergotropy stored through STIRAP charging protocol is less than half of the maximum stored ergotropy. What's more, it is interesting to find that the cdSTIRAP can significantly improve the charging power of the quantum battery in Fig. \hyperref[Fig2]{2(b)}. At the initial moment, the difference between the charging power of the two cases can be neglected. With the increase of time, the advantage of counterdiabatic driving gradually appears. The charging power gradually reaches it's maximum value at some point of time. Then, the two charging powers slowly drop as time goes on. During which the charging power of cdSTIRAP charging protocol always greater than the STIRAP one. It is worth attention that the maximum charging power increased more than four times. %In the insets of Fig. \ref{Fig2}, we display the charging process in the standard STIRAP protocol. undergo adiabatic evolution to achieve same complete charging as the cdSTIRAP protocol increase the charging time so that the quantum state in the STIRAP protocol , but the charging power is only $1\%$.
In the insets of Fig. \ref{Fig2}, we display the charging process in the standard STIRAP protocol. Obviously, to achieve the adiabatic evolution of the quantum state and make the quantum battery fully charged, one has to increase the charging time. As a result, the charging power is greatly reduced, the maximum value only $1\%$ of the cdSTIRAP protocol.

\begin{figure}
	\includegraphics[width=0.5\textwidth]{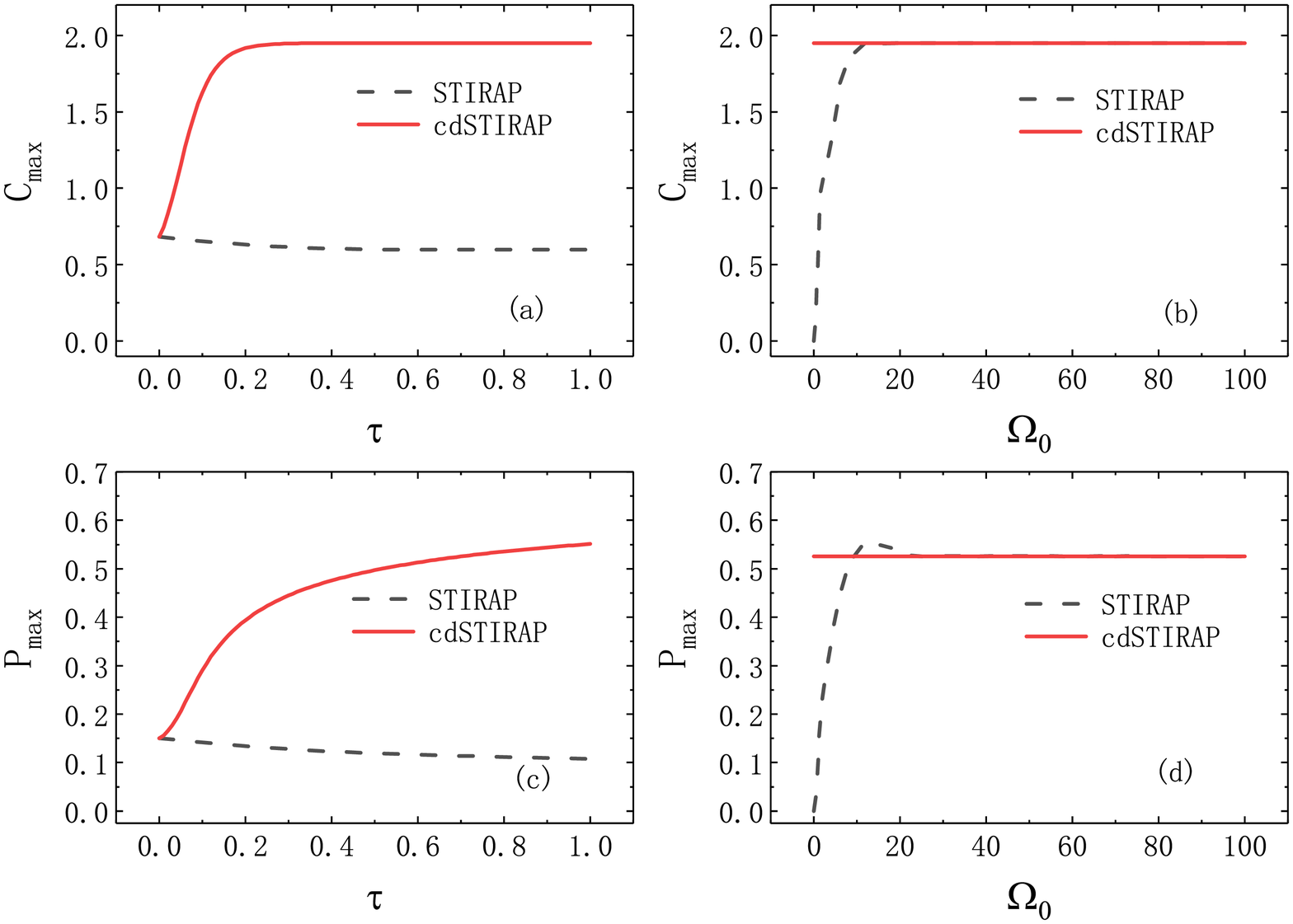}
	\caption{\label{Fig3} The maximum ergotropy $C_{max}$ as a function of (a) $\tau$ and (b) peak value $\Omega_{0}$. The dependence of the maximum charging power $P_{max}$ on (c) $\tau$ and (d) $\Omega_{0}$. For all figures, the parameter $T$ is set to $1$, and color coding, labeling and other parameters are the same as in Fig. 2.}
\end{figure}

The effect of delay $\tau$ and peak value $\Omega_{0}$ on maximum ergotropy and charging power is shown in Fig. \ref{Fig3}. Figs. \hyperref[Fig3]{3(a)} and \hyperref[Fig3]{3(c)} illustrate the dependencies of the maximum ergotropy and maximum charging power on delay time $\tau$. What is noteworthy is that the two charging protocols can receive the same result when the value of $\tau$ is taken as 0. Once the delay $\tau$ sets in, the two protocols immediately deviate from each other; the maximum ergotropy and charging power in the cdSTIRAP increase with $\tau$, whereas those in the STIRAP decrease. %Then the maximum ergotropy and charging power of one of the charging protocol are increasing while the maximum ergotropy and charging power of the other are decreasing.
% Furthermore, the STIRAP protocol can no longer inject energy to the quantum battery when $\tau$ is greater than a certain value, but the quantum battery can still be fully charged through STA, and the maximum charging power still increases slowly with $\tau$.
Fig. \hyperref[Fig3]{3(b)} and  \hyperref[Fig3]{3(d)} show the maximum ergotropy and charging power as functions of $\Omega_{0}$. We observe that the cdSTIRAP charging protocol shows good stability; for different $\Omega_{0}$ $\left(\Omega_0>0\right)$ it can always make the quantum battery fully charged, and the maximum charging power is not influenced by $\Omega_{0}$. Its advantages are quite obvious. Especially when the $\Omega_{0}$ is relatively small, the maximum ergotropy and charging power far exceed the STIRAP charging protocol.

\section{\label{4}CHARACTERS DURING DISCHARGING}
\begin{figure}
	\includegraphics[width=0.5\textwidth]{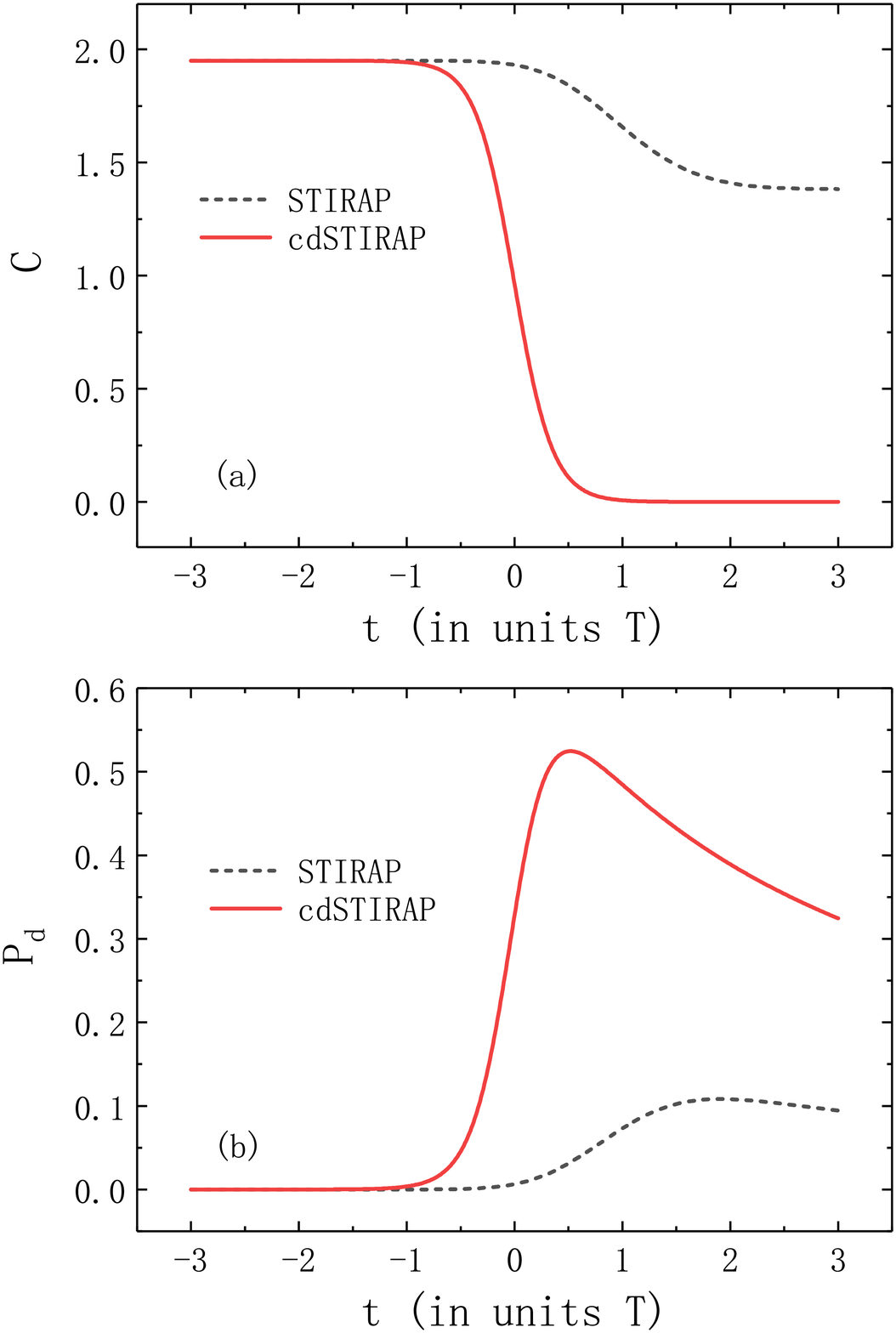}
	\caption{\label{Fig4} Time evolution of ergotropy (a) and power (b) during discharging. Color coding, labeling and parameters are the same as in Fig. 2.  }
\end{figure}
The discharging process of the quantum battery can also be realized through STA. We assume that a quantum battery is already fully charged, i.e., the three-level system is on the second excited state, and the energy is extracted by transitioning it back to the ground state. Similarly, the STA is able to accelerate this discharging process.

In STIRAP discharging protocol, energy is released by changing the sequence of pulses. More specifically, the STIRAP discharging protocol exploits a first P pulse on $|1\rangle-|2\rangle$ transition and a second S pulse on $|2\rangle-|3\rangle$ transition. That means they take the form
\begin{eqnarray}
\Omega_{p}(t)=\Omega_{0} e^{-\left(\frac{t+\tau}{T}\right)^{2}},
\Omega_{s}(t)=\Omega_{0} e^{-\left(\frac{t-\tau}{T}\right)^{2}}.
\end{eqnarray}
As for cdSTIRAP discharging protocol, to ensure the quantum battery discharge through STA, a counterdiabatic pulse is added on this basis, which satisfies
\begin{equation}
\Omega_{cd}(t)=-\frac{4\tau}{ T^{2}} \mathrm{sech}\left(\frac{4 \tau}{T^{2}}t\right).
\end{equation}

Fig. \ref{Fig4} presents the time evolution of ergotropy and power in two different discharging protocols. From fig. \hyperref[Fig4]{4(a)}, we observe that the STIRAP discharge protocol at most can only extract part of the energy stored in the quantum battery. In contrast, the cdSTIRAP discharge protocol can not only make full use of all the energy in the quantum battery, but also obviously release the energy faster. This power advantage is also illustrated in fig. \hyperref[Fig4]{4(b)}, which is more suitable for some high-power appliances.

\section{\label{5}implement}%Experimental considerations
The cdSTIRAP charging and discharging protocols described here can be implemented in different physical systems driven by optical excitations or microwave excitations, like superconducting circuit and N-V centre. Supposing we use circuit quantum electrodynamics as the experimental platform \cite{vepsalainen2019superadiabatic,doi:10.1002/qute.201900121}, the first three states of a superconducting transmon circuit constitute a three-level quantum battery. Two microwave pulses are used to realize STIRAP, and an additional two-photon microwave pulse is needed to suppress nonadiabatic excitations and thereby realize cdSTIRAP. %And charging or discharging by three microwave pulses, two of them guarantee STIRAP, while an additional two-photon microwave pulse suppress nonadiabatic excitations, which realized the cdSTIRAP.
 The negatively charged N-V centre in the diamond lattice is also an excellent experimental platform \cite{barfuss2018phase,PhysRevLett.122.090502}. It has a $S=1$ spin system which is equivalent to a three-level quantum battery in its orbital ground state. The N-V's spin sublevels are $|0\rangle$ and $|\pm 1\rangle$. It can achieve STIRAP charging or discharging protocol by driving the $|0\rangle \leftrightarrow|\pm 1\rangle$  transitions with microwave magnetic fields. In addition, applying a time-varying strain field to control the $|-1\rangle \leftrightarrow|+1\rangle$ transition, which compensate for nonadiabatic transitions and implement the cdSTIRAP.

\section{\label{6}CONCLUTIONs}
In conclusion, we have introduced the concept of a cdSTIRAP quantum battery, a three-level quantum system charged or discharged through STIRAP and %utilizes
STA based on counterdiabatic driving to accelerate charging or discharging processes. By numerical simulation of the ergotropy and power variation over time, we have confirmed the significant acceleration effect of STA on the charging and discharging process of quantum battery. The maximum charge energy can be increased approximately $3$ to $4$ times, while the maximum charge power can be increased $4$ to $5$ times. In addition, we have also studied the impact of different delay $\tau$ and peak value $\Omega_{0}$ on the quantum battery. We found that the performance of a cdSTIRAP quantum battery not vary with peak value, but raising the value of delay time would lead to a larger maximum charging power. We finally proposed that superconducting circuit or N-V centre provide a promising implementation for our cdSTIRAP quantum batteries. This study aims to provide a more efficient three-level quantum battery theoretical background in view of future experimental implementations.

\begin{acknowledgments}
The work is supported by the National Natural Science Foundation of China (Grants No. 12075193 and No. 11665020).
\end{acknowledgments}

\bibliography{Ref}

\end{document}